\documentclass[aps,pra]{revtex4-1}
\usepackage{graphicx}
\usepackage{bm}
\usepackage{dsfont}

\newcommand{\ket}[1]{\mbox{$| #1 \rangle$}}

\usepackage{amsmath,amssymb}

\usepackage[hang,normalsize]{subfigure}

\usepackage{array}
\usepackage{longtable}

\newtheorem{lemma}{Lemma}

\newcommand{\tr}{\mbox{\textbf{tr}}} 
\newcommand{\diag}{\mbox{\textbf{diag}}} 
\newcommand{\argmin}{\mbox{\textbf{argmin}}} 
\newcommand{\limi}[1]{\underset{#1}{\mbox{\textbf{lim}}}} 

\begin{document}
\preprint{}
\title{Lower Bounds for Ground States of Condensed Matter Systems}

\author{Tillmann Baumgratz}
\affiliation{Institut f\"{u}r Theoretische Physik, Albert-Einstein-Allee 11,
Universit\"{a}t Ulm, D-89069 Ulm, Germany}

\author{Martin B. Plenio}
\affiliation{Institut f\"{u}r Theoretische Physik, Albert-Einstein-Allee 11,
Universit\"{a}t Ulm, D-89069 Ulm, Germany}

\date{\today}

\begin{abstract}
Standard variational methods tend to obtain upper bounds on the ground state energy
of quantum many-body systems. Here we study a complementary method that determines
lower bounds on the ground state energy in a systematic fashion, scales polynomially 
in the system size and gives direct access to correlation functions. This is achieved 
by relaxing the positivity constraint on the density matrix and replacing it by 
positivity constraints on moment matrices, thus yielding a semi-definite programme. 
Further, the number of free parameters in the optimization problem can be reduced dramatically 
under the assumption of translational invariance. A 
novel numerical approach, principally a combination of a projected gradient
algorithm with Dykstra's algorithm, for solving the optimization problem in
a memory-efficient manner is presented and a proof of convergence for this iterative
method is given. Numerical experiments that determine lower bounds on the ground state
energies for the Ising and Heisenberg Hamiltonians confirm that the approach can be
applied to large systems, especially under the assumption of translational invariance.
\end{abstract}

\maketitle

\section{Background and Motivation}
The determination of the ground state properties of strongly interacting quantum
systems is of considerable interest. As only a very few models are exactly solvable,
numerical approximation methods are of significant importance. A key challenge
in this respect is the `curse of dimensionality', namely the exponential growth of
the Hilbert space dimension with the number of sites. Ingenious methods such as
the density matrix renormalization group (DMRG) approach \cite{White92,Schollwoeck11} can overcome this
problem as they amount to formulating a variational problem over a well-chosen class
of quantum states that can be parametrized efficiently, i.e. with a polynomial number
of parameters \cite{RommerOstlund1995}. As the variation is over a subclass of all
possible quantum states, variational methods of this type provide upper
bounds on the ground state energy. Assessing the quality of the so obtained upper
bounds on the ground state energy is then difficult. Hence there is considerable
interest in the development of efficient methods for the determination of lower bounds
on the ground state energy. One straightforward approach in this direction consists
in the determination of Anderson bounds \cite{AndersonLower}, i.e. the decomposition of a
Hamiltonian $H=\sum_k H_k$ and the subsequent determination of the ground state energy
of each $H_k$ whose sum then yields a lower bound on the ground state energy of $H$.
While useful, this approach does not scale well with the system size as it remains
exponential in the size of the support of the $H_k$. The application of DMRG to the $H_k$
may overcome the scaling issue but amounts to finding an `upper bound to a lower bound'
and does not resolve the principal issue with DMRG.

As early as $1940$, however, it was suggested to avoid the use of wave functions altogether
and rather consider reduced density matrices together with conditions that these arise from
a valid quantum state \cite{husimi40} -- the so-called N-representability problem
\cite{Coleman,Garrod}. Varying over the reduced density matrices that are compatible
with a valid physical state then allows for the determination of the ground state energy.
The exact ground state energy can be obtained if all conditions for the N-representability
problem are known. If only partial conditions are specified, then one minimizes over too
large a set of reduced density matrices and one obtains a lower bound on the ground state
energy. The analytical treatment of this problem is highly challenging and at the time
numerical solutions were hard to come by. Further progress was made by Mazziotti
\cite{mazziotti01,mazziotti07},  who pointed out that the problem can be formulated
as a semi-definite programme and use can be made of the guaranteed and certified
convergence of the primal-dual interior-point algorithm for semi-definite programming
\cite{boyd09,vandenberghe96}. A general mathematical framework for this type of optimization 
problems and applications was given in \cite{Pironio10}. Unfortunately, the primal-dual 
interior-point algorithm and related methods do come with a considerable overhead in 
computation time and, crucially, memory that is making it hard to scale to very large systems. 
Lately, new algorithms to address the optimization problems of finding lower bounds on ground state 
energies have been developed \cite{mazziotti04,mazziotti11} and considerable improvements towards 
the interior-point algorithms have been demonstrated. 

In this work, we are exploring further the idea of relaxations of the ground state
energy finding problem that allow us to determine lower bounds on the ground state
energy in an efficient manner. To this end we formulate the problem, introduce a
number of simplifications including the explicit treatment of translation invariance
and symmetries and point out additional potential routes for numerical simplifications.
In addition, we present a novel algorithm, principally a combination of a projected
gradient algorithm with Dykstra's algorithm \cite{Dattorro,deutsch01,Baumgratz}, for solving the
resulting optimization problem in a memory-efficient manner and prove its convergence. 
Then, in order to demonstrate the capabilities of this approach, we present numerical
experiments that determine lower bounds on the ground state energies for two one-dimensional (1D) lattice models, 
namely the Ising and Heisenberg Hamiltonian, which confirm that the approach can be applied for large systems
especially when implementing translational invariance. Note that although we restrict our numerical analysis to 1D
lattice models it has been demonstrated that the variational determination of ground state energies considering only constraints on 
reduced density matrices applies to higher spatial dimensions \cite{mazziotti10} and to non-periodic systems such as molecules 
in quantum chemistry \cite{mazziotti04} as well. We complete this work with a discussion and outlook.\vspace*{-0.5cm}

\section{Basics}
\label{sec:basics}
We begin by explaining the general ideas underlying the method for the determination of
lower bounds on ground state energies following \cite{mazziotti01} and then specialize to the case of fermionic systems, for which later sections will provide numerical 
examples. Analogous ideas can also be implemented for bosonic systems or general spin-systems.

\subsection{Generalities}
The determination of the ground state energy $E_{0}$ of a many-body system composed of
$N$ particles and described by the Hamilton operator $H$ can be formulated as the minimization
problem
\begin{equation}
    E_{0}=\min \left\{\tr\left[H\rho\right]\, \vert \, \rho\geq 0,\, \tr\left[\rho\right]=1\right\}. 
    \label{eqn:originaloptimizationproblem}
\end{equation}
This approach is pushed to its limits rather rapidly as it uses the positivity
constraint on $\rho$, which grows dramatically with the system size, e.g. for particles 
with spin-$1/2$ or spinless fermions $\rho$ is a $2^N \times 2^N$ matrix. Hence, this problem cannot be
solved efficiently due to the exponentially increasing parameter space. One technique to
deal with this difficulty is to parametrize a specific family of candidates $\ket{\phi\left(\vec{\alpha}\right)}$
for the ground state by a set of parameters $\vec{\alpha}$. One then minimizes the
expectation value of the Hamiltonian with respect to this family of states. As this
family of states does not encompass the entire state space, this method will find a
state $\ket{\phi_{0}}$ which is optimal in $\left\{\ket{\phi\left(\vec{\alpha}\right)}\right\}_{\vec{\alpha}}$
and which yields an upper bound on the true ground state energy. This is the route taken
by a variety of variational methods including, for example, DMRG \cite{White92,Schollwoeck11,RommerOstlund1995}.

To overcome the scaling issues and to provide lower rather than upper bounds on the
ground state energy, we need to relax the positivity constraint $\rho\ge 0$, replacing
it by a weaker set of constraints which scale only polynomially in the system's size 
\cite{mazziotti01,Pironio10}. This enlarges the class of states over which one
varies the energy and hence yields lower bounds on the ground state energy. 
To this end we use the fact that
\begin{equation}
    \rho \geq 0 \Rightarrow \tr\left[\mathcal{O}^{\dagger}\mathcal{O}\rho\right] \geq 0
\end{equation}
for all operators $\mathcal{O}$. Demanding only
\begin{equation}
    \langle \mathcal{O}^{\dagger}\mathcal{O}\rangle =
    \tr\left[\mathcal{O}^{\dagger}\mathcal{O}\rho\right] \geq 0
    \label{eqn:relaxedconstraint}
\end{equation}
for some operator $\mathcal{O}$ does not, in general, imply $\rho \geq 0$. For some subset
of operators $\{O_k\}_{k}$, we can then define $\mathcal{O} = \sum_k \alpha_k O_k$ and the
requirement that $\langle \mathcal{O}^{\dagger}\mathcal{O}\rangle\ge 0$ for all choices of $\alpha_k$
is then equivalent to the positivity of the matrix $X$ with components
$X_{kl}=\langle O_{k}^{\dagger} O_{l}\rangle$, imposing a particular N-representability condition. 
Hence, replacing in the original optimization
problem eq. (\ref{eqn:originaloptimizationproblem}) the condition $\rho\ge 0$ with
the weaker condition $X \geq 0$ and omitting the normalization $\tr[\rho]=1$, we find that
\begin{equation}
    E_{0} \ge \min \left\{\tr\left[H\rho\right]\, \vert \, X \ge 0\right\}.
    \label{eqn:originaloptimizationproblemrelaxed}
\end{equation}
The present formulation still contains the density operator $\rho$ explicitly
in the computation of $\tr\left[H\rho\right]$. Judicious choice of the set
$\{O_k\}_{k}$ alleviates this shortcoming. Indeed, writing $H=\sum_{kl} h_{kl} H_{kl}$, 
where $h_{kl}\in\mathbb{C}$ are the coefficients of the operators $H_{kl}$ in the Hamiltonian $H$, 
one ensures that the set $\{O_k\}_{k}$ contains all operators such that for all $k,l$ the expectation value 
$\langle H_{kl} \rangle$ is in $\{\langle O_{k}^{\dagger}O_{l}\rangle\}_{k,l}$. Note, furthermore, that
the entries of $X$ will generally not be independent but suffer some linear
equality constraints, for example due to (anti)-commutation relations between operators.
Denoting by $\mathcal{C}$ the set of all Hermitian matrices that satisfy
$X\ge 0$ and any other linear constraints, the energy minimization problem can 
be written as \cite{mazziotti01,mazziotti07} 
\begin{equation}
    E_{0} \ge \min \left\{ \sum_{kl} h_{kl} X_{kl}\, \vert \, X \in \mathcal{C}\right\},
    \label{eqn:originaloptimizationproblemrelaxedlinear}
\end{equation}
which is evidently a semi-definite programme \cite{boyd09} and efficiently solvable
as long as the set $\{O_k\}_{k}$ contains a number of operators that scale polynomially
in system size. The decomposition of $H$ into $H_{kl}$, and hence the choice of the set
$\{O_k\}_k$, is not unique and the optimal choice is not obvious. From a practical point
of view however, the systematic choice presented in the remainder of the paper leading 
to the so-called p-positivity conditions \cite{mazziotti01,mazziotti07} does perhaps provide the most straightforward formulation.

\subsection{Bosons and Fermions}
For concreteness, let us now consider the class of bosonic or fermionic operators defined by
\begin{eqnarray}
    \mathcal{O}&=&\sum_{k}\delta_{k} a_{k} + \sum_{k} \epsilon_{k} a_{k}^{\dagger} 
    +\sum_{kl}\alpha_{kl} a_{k}	a_{l} + \sum_{kl}\beta_{kl}a_{k}^{\dagger}a_{l} + \sum_{kl}\gamma_{kl}a_{k}^{\dagger}a_{l}^{\dagger},
    \label{eqn:definitiongeneraloperator}
\end{eqnarray}
where $\delta_{k},\epsilon_{k},\alpha_{kl},\beta_{kl},\gamma_{kl}\in\mathbb{C}$ are
arbitrary complex variables. The N-representability conditions arising from this class of operators are called 1-positivity and 2-positivity conditions, 
which is a terminology used to reveal the powers of the creation and annihilation operators $a_{i}^{\dagger}$ and $a_{i}$ in $\mathcal{O}$. In general, allowing polynomials of 
order p in the second-quantized operators leads to restrictions called p-positivity conditions \cite{mazziotti01,mazziotti07}. 
For the particular choice of operators of the order of 2, requiring the positivity of $\langle \mathcal{O}^{\dagger}\mathcal{O} \rangle$ is equivalent to
\begin{equation}
    X=\begin{pmatrix}  T & U^{\dagger} & A^{\dagger} & B^{\dagger} & C^{\dagger} \\
			U & S & D^{\dagger} & E^{\dagger} & F^{\dagger} \\
			A & D & M & G^{\dagger} & H^{\dagger} \\
			B & E & G & R & I^{\dagger} \\
			C & F & H & I & Q \end{pmatrix}  \geq 0,
    \label{eqn:definitionmatrixC}
\end{equation}
where the $N \times N$ matrices comprising the second moments are defined as follows:
\begin{eqnarray}
    T_{k,l} &=& \langle a_{k}^{\dagger} a_{l}\rangle, \nonumber\\
    S_{k,l} &=& \langle a_{k}a_{l}^{\dagger} \rangle, \nonumber\\
    U_{k,l} &=& \langle a_{k}a_{l} \rangle
\end{eqnarray}
and for the third and fourth moments, we find the $N^2 \times N$ and $N^2 \times N^2$ matrices
\begin{eqnarray}
A_{kl,m}&=\langle a_{k}^{\dagger}a_{l}^{\dagger} a_{m} \rangle,\hspace{1cm} M_{kl,mn}&=\langle a_{k}^{\dagger}a_{l}^{\dagger} a_{n}a_{m} \rangle, \nonumber\\
B_{kl,m}&=\langle a_{k}^{\dagger}a_{l}a_{m} \rangle,\hspace{1cm} G_{kl,mn}&= \langle a_{k}^{\dagger} a_{l}a_{n}a_{m} \rangle, \nonumber\\
C_{kl,m}&=\langle a_{k}a_{l}a_{m} \rangle,\hspace{1cm} H_{kl,mn}&= \langle a_{k}a_{l}a_{n}a_{m} \rangle, \\
D_{kl,m}&=\langle a_{k}^{\dagger}a_{l}^{\dagger}a_{m}^{\dagger} \rangle,\hspace{1cm} R_{kl,mn}&= \langle a_{k}^{\dagger}a_{l}a_{n}^{\dagger}a_{m} \rangle, \nonumber\\
E_{kl,m}&= \langle a_{k}^{\dagger}a_{l}a_{m}^{\dagger} \rangle,\hspace{1cm} I_{kl,mn} &= \langle a_{k}a_{l}a_{n}^{\dagger}a_{m} \rangle,\nonumber\\
F_{kl,m}&= \langle a_{k}a_{l}a_{m}^{\dagger} \rangle,\hspace{1cm} Q_{kl,mn}&= \langle a_{k}a_{l}a_{n}^{\dagger}a_{m}^{\dagger} \rangle. \nonumber
\end{eqnarray}
In addition to the positivity constraint on the $2N+3N^2 \times 2N+3N^2$ Hermitian matrix $X\in\mathcal{H}$, where $\mathcal{H}$ denotes the vector space of all Hermitian matrices with
appropriate dimension, there are linear constraints relating its entries. In fact, the fermionic anti-commutator relations
\begin{equation}
\left\{a_{k},a_{l}\right\} = \left\{a_{k}^{\dagger},a_{l}^{\dagger}\right\} = 0, \quad \left\{a_{k},a_{l}^{\dagger}\right\}=\delta_{k,l}
\end{equation}
or the bosonic commutator relations
\begin{equation}
\left[a_{k},a_{l}\right] = \left[a_{k}^{\dagger},a_{l}^{\dagger}\right] = 0, \quad \left[a_{k},a_{l}^{\dagger}\right]=\delta_{k,l}
\end{equation}
force some entries of $X$ to be equal, to differ by sign or to be a linear function of others.

Now, let $\mathcal{C}$ be the set of all positive semi-definite Hermitian matrices of the form eq.
\eqref{eqn:definitionmatrixC} and satisfying the additional requirement that the
entries of $X$ obey the constraints given by the commutation or anti-commutation
relations. Further, let $E\left(X\right): \mathcal{H}
\rightarrow \mathbb{R}$ be a function expressing the expectation value of a Hamiltonian
with up to 4-mode terms, i.e. $E\left(X\right)=\langle H \rangle$. Then, the original optimization problem eq. \eqref{eqn:originaloptimizationproblem} can be replaced by
\begin{equation}
    E_{0}\geq \min \left\{E\left(X\right) \, \vert \, X \in \mathcal{C} \right\},
\label{eqn:relaxedoptimizationproblem}
\end{equation}
where the relaxed optimization problem scales polynomially in the system size $N$. Note that
it is straightforward to extend this formulation to higher moments, e.g. 3-positivity conditions. In the following, we are going to
present scenarios and strategies where the number of free variables can be decreased considerably.

\subsection{Consequences of Superselection Rules and Particle Number Conservation for Fermions}
For the remainder we will consider fermionic systems. In that
case, superselection rules require that the Hamiltonian commutes with the parity
operator given by $P=\prod_{k} (1-2a^{\dagger}_{k}a_{k})$. As a consequence,
only Hamiltonians comprising terms with an even number of creation and annihilation
operators are physically allowed. This in turn ensures that only expectation values
of products of an even number of creation and annihilation operators are non-vanishing.
Hence, we find that we can restrict $X$ to take the form
\begin{equation}
X=\begin{pmatrix}  T & U^{\dagger} & 0 & 0 & 0 \\
			U & S & 0 & 0 & 0 \\
			0 & 0 & M & G^{\dagger} & H^{\dagger} \\
			0 & 0 & G & R & I^{\dagger} \\
			0 & 0 & H & I & Q \end{pmatrix}
\label{eqn:matrixXaftersuperselectionrules}
\end{equation}
and to satisfy the constraints dictated by the fermionic anti-commutator relations.
See appendix \ref{app:constraints} for a list of constraints on the entries for a formulation for
up to fourth moments in the fermionic case.

If the Hamiltonian conserves the particle number, a further simplification
ensues. In this case, each term in the Hamiltonian will consist of an equal
number of annihilation and creation operators. Hence, the Hamiltonian is invariant
under the transformation $a_k \rightarrow a_k e^{i\phi}$ and $a_k^{\dagger} \rightarrow
a_k^{\dagger} e^{-i\phi}$ and we find that expectation values vanish unless they
concern operators with the same number of annihilation and creation operators so
that $X$ can be assumed to take the form
\begin{equation}
    X=\begin{pmatrix} T & 0 & 0 & 0 & 0 \\
                      0 & S & 0 & 0 & 0 \\
                      0 & 0 & M & 0 & 0  \\
                      0 & 0 & 0 & R & 0  \\
                      0 & 0 & 0 & 0 & Q
    \end{pmatrix}. 
\label{eqn:matrixforfermionsnumberconservation}
\end{equation}
Note that the elements of the matrices $T,S$ are coefficients of the one-electron reduced density matrix (1-RDM), while the matrices $M,R,Q$ are representations of the two-electron reduced density matrix (2-RDM) \cite{mazziotti07}. Restricting matrices of the form eq. \eqref{eqn:matrixforfermionsnumberconservation} to be positive semi-definite is equivalent with the positivity of the individual matrices $T,S$ and $M,R,Q$. This constraint on the matrices containing the second moments defines the 1-positivity conditions, while the positivity of the matrices containing the fourth moments generates the 2-positivity conditions \cite{mazziotti01,mazziotti07}. 
For concreteness we list the matrices comprising the constraints on a system composed of two spinless fermions and the additional assumption of particle number conservation in appendix 
\ref{app:constraints}.

\subsection{Exploiting Translational Invariance}
So far we considered optimization problems for up to 2-positivity conditions which reduced the number of free variables to scale as $N^{4}$. 
Apart from superselection rules and the particle number conservation, no specific assumption on the structure of the underlying Hamiltonian has been exploited such that 
in principle this procedure is applicable to a wide range of different systems, e.g. quantum chemistry calculations \cite{mazziotti04} or higher spacial 
dimensions \cite{mazziotti10}. Since our intention is to apply this method to large lattice models, we now 
discuss a further simplification of the system to decrease the number of parameters, namely the translational invariance.
Translation invariance with periodic boundary conditions of the physical system reduces the 
number of free variables considerably. In fact for the second moments, i.e. for the matrices
$T,S$ and $U$, this can be exploited immediately as these matrices then obey $T_{i,j}=t_{i-j}$, etc.
Consequently, for translational invariant systems with periodic boundary conditions, the matrices
comprising the second moments become circulant matrices \cite{horn91} and can therefore be
diagonalized by a discrete Fourier transform
\begin{equation}
V_{k,l}=\frac{1}{\sqrt{N}} e^{-2\pi i\left(k-1\right)\left(l-1\right)/N}
\label{eqn:discretefourierunitary}
\end{equation}
for $k,l=1,\ldots,N$.
Reducing the number of variables for the fourth moments is not so transparent as for the second moments, but manageable. To do so, note that the ordering of the entries of the matrices
$M,G,H,R,I$ and $Q$ is not fixed but can be chosen arbitrarily. Remember that the indices
of these matrices are of the form $\left[\left(k,l\right),\left(m,n\right)\right]$ for $k,l,m,n=1,\ldots,N$. Rearranging the two-digit indices labeling the rows and columns in the following way:
\begin{equation}
\begin{split}
\left(k,l\right)=&\left\{ \left(1,1\right), \left(2,2\right), \ldots,\left( N,N\right), \right. \\
&\left(1,2\right), \left(2,3\right), \ldots,\left( N,1\right), \\
&\left(1,3\right), \left(2,4\right), \ldots, \left(N,2\right), \\
& \ldots  \\
& \left(1,N\right), \left(2,1\right), \ldots,\left( N,N-1\right)\left.\right\},
\end{split}
\end{equation}
the matrices comprising the fourth moments decompose in a block structure where each block can be diagonalized by the unitary matrix $V_{k,l}$ given above. For example, we find for the
matrix $M$
\begin{equation}
M=U D_{M} U^{\dagger}, 
\end{equation}
where $U=\diag\left(V,V,\ldots,V\right)$ with $V$ given by eq. \eqref{eqn:discretefourierunitary} and
\begin{equation}
D_{M}=\begin{pmatrix} D^{1,1} & D^{1,2} & \ldots & D^{1,N} \\
				 \vdots & \vdots & \ddots & \vdots \\
				 D^{N,1} & D^{N,2} & \ldots & D^{N,N} \end{pmatrix}
\end{equation}
with diagonal matrices $D^{i,j}$, $i,j=1,\ldots,N$. Hence, for translational invariant systems the matrices contained in set $\mathcal{C}$ will have the following form:
\begin{equation}
X=\begin{pmatrix}  D_{T} & D_{U^{\dagger}} & 0 & 0 & 0 \\
			D_{U} & D_{S} & 0 & 0 & 0 \\
			0 & 0 & D_{M} & D_{G^{\dagger}} & D_{H^{\dagger}} \\
			0 & 0 & D_{G} & D_{R} & D_{I^{\dagger}} \\
			0 & 0 & D_{H} & D_{I} & D_{Q} \end{pmatrix},
\end{equation}
where the matrices denoting the second moments in the upper left block are diagonal matrices
and the matrices considering the fourth moments have a block diagonal structure as described
above. These considerations reduce the complexity of the constraints considerably
and hence make the numerical study feasible for large systems.

\subsection{Varying System Size for Higher Moments}
As the treatment of the fourth-order correlations is computationally costly, it may in some cases be advantageous to consider fourth-order constraints on a smaller lattice, while the 
computationally cheaper second-order constraints are treated on the full lattice. In particular, while $T,U$ and $S$ are $N\times N$ matrices, $M,G,H,R,I$ and $Q$ are $N^2\times N^2$
matrices leading to the unfavourable scaling. To overcome this issue, it might therefore 
be beneficial to restrict higher moments to smaller subsystems; that is, for the
second moments one would consider system size $N_2$, while for fourth moments one
would consider $N_{4}$ where $N_{2}\gg N_{4}$. Reducing the support of the matrices
$M,G,H,R,I$ and $Q$, on the one hand, will lead to disproportionate savings in memory and computation
time and, on the other hand, will certainly lead to a decrease of the lower bounds.


\section{The Projected Gradient Algorithm}
So far we have described how to relax the original optimization problem eq.~\eqref{eqn:originaloptimizationproblem} and 
discussed how to include symmetries to reduce the number of free variables. Now we are
going to describe an iterative algorithm for addressing this minimization problem. Recall
the minimization problem we have to face, that is,
\begin{equation}
    \begin{array}{ll}
    \mbox{minimize} & E\left(X\right)  \\
    \mbox{subject to} & X \in \mathcal{C} \\
    \end{array}
    \label{eqn:relaxedminimizationproblem}
\end{equation}
where again the function $E\left(X\right):\mathcal{H}\rightarrow \mathbb{R}$ expresses
$\langle H \rangle$ and the set $\mathcal{C}$ consists of all positive semi-definite
Hermitian matrices $X$ fulfilling the constraints dictated by the (anti)-commutator relations and potential other symmetries of the system.
Note that the function $E\left(X\right)$ is indeed linear in the elements of matrix $X$ and
hence could be written as
\begin{equation}
    E\left(X\right) = \tr\left[G \cdot X\right] +c,
\end{equation}
where $G$ is a Hermitian matrix and $c\in\mathbb{R}$. Furthermore, the set $\mathcal{C}$ is
the intersection of the convex cone consisting of positive semi-definite matrices and the
affine set defined by the linear constraints on the entries. Hence $\mathcal{C}$
is convex, too~\cite{boyd09}. It is well known that this optimization problem is in fact
a semi-definite programme and as such can be solved efficiently. Standard primal-dual
interior point methods for semi-definite programming come with a considerable overhead
in memory and time which makes their application to large systems challenging. Recently, two algorithms 
addressing the problem of ground state energy estimation were developed which scale better than the conventional 
methods \cite{mazziotti04,mazziotti11}. Here, we continue the research for efficient algorithms and 
formulate an alternative numerical method whose efficacy we demonstrate in later sections.

\begin{figure}
\begin{center}
\includegraphics[width=0.40\textwidth]{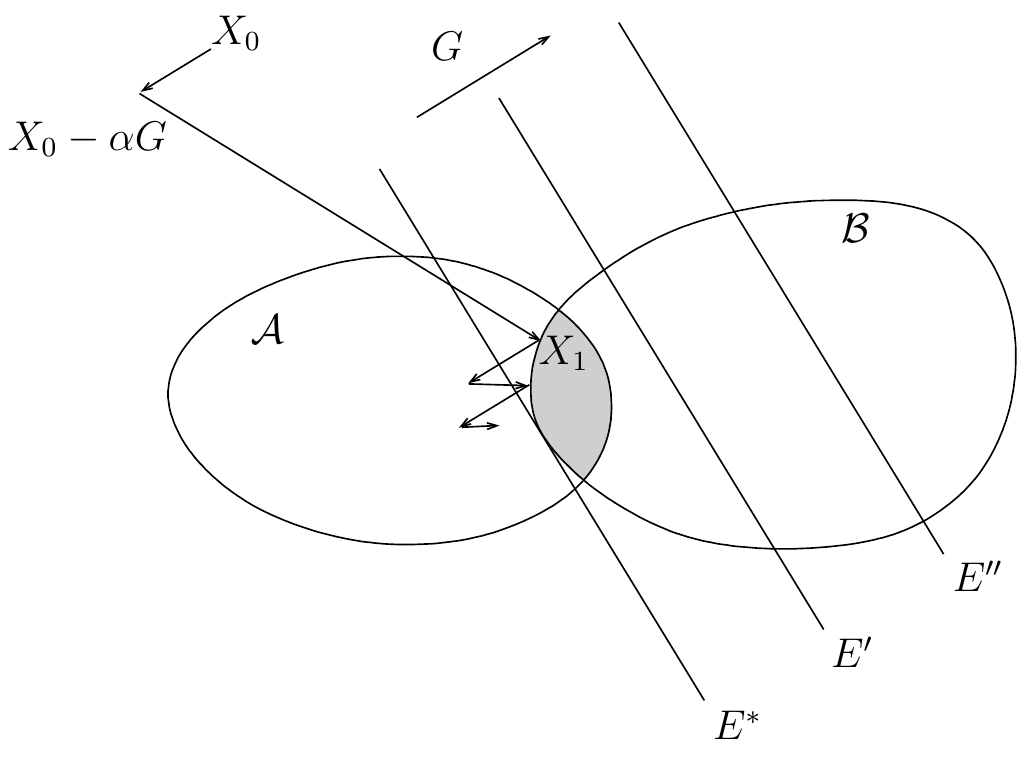}\vspace{-0.7cm}
\end{center}
\caption{Illustration of the gradient projection algorithm \eqref{eqn:gradientprojectionalgorithm}. The convex sets $\mathcal{A}$ and $\mathcal{B}$ have a nonempty intersection, which is
shown shaded. The straight parallel lines clarify the level sets of the function $E\left(X\right)$ such that $E^{*} < E^{\prime} < E^{\prime \prime}$, where $E^{*}$ is the minimum value the
function attains on the intersection. The projections are represented by arrows to demonstrate their directions. Note that these projections are computed using Dykstra's algorithm, which
itself uses the projections onto $\mathcal{A}$ and $\mathcal{B}$. The algorithm is initialized with the matrix $X_{0}$.   }
\label{fig:algorithm}
\end{figure}

To address the issues described above, we suggest to solve the minimization problem by a projected gradient
algorithm \cite{calamai87}. Since this algorithm relies heavily on projections onto convex
sets, we need to define the latter. Let $\mathcal{H}$ be a vector space and $\mathcal{A}\in \mathcal{H}$
be a convex set. The projection of $X\in \mathcal{H}$ onto $\mathcal{A}$ is defined by
\begin{equation}
    P_{\mathcal{A}}\left(X\right)=\argmin\left\{ || X-Y ||_{F} \, \vert \, Y\in \mathcal{A}\right\},
    \label{eqn:definitionprojectionontoset}
\end{equation}
where $|| \cdot ||_{F}:\mathcal{H}\rightarrow \mathbb{R}$ denotes the Frobenius or Hilbert-Schmidt
norm. As the name suggests, the projection determines the nearest point in the convex set
$\mathcal{A}$ with respect to a given norm. With this, we claim that the solution of problem eq. \eqref{eqn:relaxedminimizationproblem} could be found by iterating
\begin{equation}
    X_{k+1}=P_{\mathcal{C}}\left(X_{k} - \alpha G\right),
    \label{eqn:gradientprojectionalgorithm}
\end{equation}
where the algorithm is initialized by a Hermitian matrix $X_{0}\in\mathcal{H}$ and $\alpha >0$ is
an arbitrary real and positive number. Furthermore, $P_{\mathcal{C}}:\mathcal{H}\rightarrow \mathcal{C}$ denotes the projection onto set $\mathcal{C}$, i.e. the intersection of the positive 
semi-definite Hermitian matrices and the set given by the linear constraints. The scheme of this algorithm is visualized in figure \ref{fig:algorithm}.

We find that the following lemma holds.
\begin{lemma}
Let $E:\mathcal{H}\rightarrow \mathbb{R}$ be a linear function of the form $E\left(X\right)=\langle G, X \rangle + c$ where $G\in \mathcal{H}^{n\times n}$ and $c\in\mathbb{R}$
is a constant. Assume that the function $E\left(X\right)$ is lower bounded on the closed convex set $\mathcal{C}$, i.e. there is an $\bar{X}\in\mathcal{C}$ such that $E\left(\bar{X}\right)
\leq E\left(X\right)$ for all $X\in\mathcal{C}$. Let $E_{k}=E\left(X_{k}\right)$ be the sequence determined by the iterations of the algorithm eq.~\eqref{eqn:gradientprojectionalgorithm}. Then
$E_{k}=E\left(X_{k}\right)\rightarrow E^{*}$ where $E^{*}\leq E\left(X\right)$ for all $X\in\mathcal{C}$.
\label{lemma:gradientprojection}
\end{lemma}
A detailed proof of this lemma is presented in appendix \ref{app:proofconvergence}.

So far we have shifted the problem of minimizing the function $E\left(X\right)$ with
respect to the convex set $\mathcal{C}$ to the problem of computing the projection onto
$\mathcal{C}$. As pointed out above, this set is the intersection of two convex
sets. This complicates the computation of the projection because in principle one has to
solve the optimization problem eq.~\eqref{eqn:definitionprojectionontoset} in each step of
the algorithm. The most obvious method at hand to overcome this difficulty is to use Dykstra's
algorithm \cite{deutsch01,birgin05}. Dykstra's algorithm is an elaborate technique to
compute the projection of $X$ onto the intersection of several convex sets by means of
modified iterative projections onto the individual sets. To apply this procedure to the
computation onto $\mathcal{C}$ we need to determine the projections onto the set of
positive semi-definite Hermitian matrices and onto the affine set given by the constraints
imposed by the canonical (anti)-commutation relations and other symmetries. The former can be implemented
straightforwardly. For the latter we were able to determine formulae for the submatrices in 
eq.~\eqref{eqn:matrixXaftersuperselectionrules} and thus gain a method to compute $P_{\mathcal{C}}: \mathcal{H}\rightarrow \mathcal{C}$ efficiently. Note that since $\mathcal{C}$ is convex,
$P_{\mathcal{C}}\left(X\right)$ is unique for all $X\in\mathcal{H}$ \cite{deutsch01}. 
Further, the proposed algorithm finds a lower bound on the ground state energy by only considering boundary points of the 
feasible set and hence is slightly related to the algorithm proposed in \cite{mazziotti11}.

There are two parameters defining the accuracy of the estimated lower bound. In principle
the algorithm \eqref{eqn:gradientprojectionalgorithm} converges for perfect projections $P_{\mathcal{C}}\left(X\right)$ to a matrix $X\in\mathcal{C}$ minimizing the function
$E\left(X\right)$ in the limit $k\rightarrow \infty$, where $k$ is the iteration number.
As for all iterative algorithms we need to introduce stopping criteria, on the one hand,
for the projected gradient algorithm and, on the other hand, for Dykstra's algorithm. For the latter
we use the robust stopping criterion introduced in \cite{birgin05} and determined by a small number
${\tau_{Dykstra}\in\mathbb{R}}$. For the former one can exploit the fact that the dual problem of the
optimization problem eq. \eqref{eqn:relaxedminimizationproblem}, which we identify as the primal problem, can be solved
using the projected gradient algorithm, too. Note that the primal problem can be written as
\begin{equation}
\begin{array}{ll}
\mbox{minimize} & \vec{g}^{\dagger}\vec{x}  \\
\mbox{subject to} & F_{0}+\sum_{i}x_{i} F_{i} \geq 0, \\
\end{array}
\label{eqn:primalproblem}
\end{equation}
where we identify the matrices $G$ and $X$ with the vectors $\vec{g}$ and $\vec{x}$ such that $E\left(X\right)-c=\langle G, X \rangle =\vec{g}^{\dagger}\vec{x}$.
The Hermitian matrices $F_{0},\left\{F_{i}\right\}_{i}$ comprise the linear constraints dictated by the fermionic anti-commutator relations such that the matrix
$F_{0}+\sum_{i}x_{i} F_{i}$ is an element of the affine set for all $x_{i}\in\mathbb{R}$. The dual problem of eq. \eqref{eqn:primalproblem} is then for
the dual variable $Z\in\mathcal{H}$ \cite{vandenberghe96}
\begin{equation}
\begin{array}{ll}
\mbox{maximize} & - \langle F_{0} , Z \rangle   \\
\mbox{subject to} & g_{i}=\langle F_{i} , Z \rangle \\
			  & Z \geq 0.
\end{array}
\label{eqn:dualproblem}
\end{equation}
Note that the constraints of the dual problem force the solution $Z^{*}$ to lie both in the positive semi-definite cone and in the affine set given by the linear equations.
The intersection of these two convex sets is again convex and the dual problem becomes
\begin{equation}
\begin{array}{ll}
\mbox{minimize} & \langle F_{0} , Z \rangle  \\
\mbox{subject to} & Z \in \mathcal{D}, \\
\end{array}
\label{eqn:dualproblem2}
\end{equation}
where $\mathcal{D}$ denotes the intersection of the positive semi-definite cone with the set given by the
linear constraints $g_{i}=\langle F_{i} , Z \rangle$. Finally, lemma \ref{lemma:gradientprojection} shows that the projected gradient algorithm can be applied to solve the dual problem as well.

The benefit of the formulation as a primal-dual problem is that the solution of the dual problem is a lower bound on the solution of the primal problem. 
Hence, solving the primal and dual problems simultaneously using the projected gradient algorithm provides us with an ingenious certificate for
the accuracy of the solution. Consequently, one way to introduce a stopping criterion for the projected gradient algorithm is to initiate a small quantity $\tau_{primal-dual}$ and stop the
computations when
\begin{equation}
E\left(X_{k}\right) - H\left(Z_{k}\right)  \leq \tau_{primal-dual},
\label{eqn:dualitygap}
\end{equation}
where $H\left(Z\right)=-\langle F_{0} , Z \rangle +c$ is basically the dual function up to a constant. Note that eq. \eqref{eqn:dualitygap} is equivalent to the duality gap of the primal and dual
problem and that for strictly feasible primal problems, i.e. for optimization problems of the form eq. \eqref{eqn:primalproblem} for which there exists an $x$ such that $ F_{0}+\sum_{i}x_{i} F_{i}
> 0$, the duality gap is zero if evaluated at optimal points~\cite{vandenberghe96}. For the optimization problem eq. \eqref{eqn:relaxedminimizationproblem}, one can easily find a point in the
feasible set which is strictly feasible. Consequently, we have $E\left(X^{*}\right) - H\left(Z^{*}\right) =0$, where $X^{*}$ and $Z^{*}$ are 
solutions of the primal and dual problems, respectively, and hence eq. \eqref{eqn:dualitygap} suits perfectly as a stopping criterion. Due to the numerical cost to solve the primal and dual
problems simultaneously, we decide on using a different stopping criterion for the projected gradient algorithm. We stop when the improvement per iteration step, i.e.
\begin{equation}
    E\left(X_{k+1}\right)-E\left(X_{k}\right) \leq \tau_{E},
\end{equation}
falls below a predetermined small number $\tau_{E}\in\mathbb{R}$. We implemented both routines for systems considering only second moments and numerical experiments
suggest that for a sufficiently small number $\tau_{E}$ the latter stopping criterion yields results that are as good as in the primal-dual formalism.

Another degree of freedom in the formulation of the algorithm is the step length $\alpha$. Numerical examples propose that for the first step in the direction of the negative gradient $\alpha$
should be chosen to be large. For the remaining steps the computational cost could be reduced by choosing a small but appropriate value of $\alpha$ \cite{remarkalgorithm}. 

\section{Applications}
In the remainder of this work, we provide a number of numerical examples that demonstrate
that the proposed method can be applied to rather large systems and is capable of extracting
accurate information concerning the ground state energy as well as the correlation functions. 
The purpose of what follows is not to improve lower bounds on unknown ground state energies, but 
rather to analyse the behaviour of the proposed algorithm for finite but large lattice models.

\subsection{Second Moments and the Ising Model}
We begin our exposition of the numerical application of the algorithm with a
simple, exactly solvable model for which the relaxation of the energy minimization problem has been analysed in great detail exploiting 
different numerical methods in \cite{Mazziotti09}. 
The Hamiltonian we are considering as a first application of the projected gradient algorithm is given by 
\begin{eqnarray}
    H &=& j_{c} \sum_{k=1}^{N-1}\left( a_{k}^{\dagger}a_{k+1} + a^{\dagger}_{k+1}a_{k} \right)
    - j_{c} \left(a^{\dagger}_{N}a_{1} + a_{1}^{\dagger}a_{N} \right) \nonumber \\
    && + j_{c} \sum_{k=1}^{N-1}\left(a^{\dagger}_{k}a_{k+1}^{\dagger} + a_{k+1}a_{k} \right) - j_{c} \left(a^{\dagger}_{N}a_{1}^{\dagger} + a_{1}a_{N} \right)
    \nonumber \\
    && -hN + 2h \sum_{k=1}^{N}a_{k}^{\dagger}a_{k},
    \label{eqn:isinghamiltonian}
\end{eqnarray}
where the operators describe spinless fermions. The ground state of this model coincides with that of the Ising model in a transverse magnetic field with periodic boundary
conditions
\begin{equation}
    H = j_c \sum_{k=1}^N \sigma_k^{x}\sigma_{k+1}^{x} - h\sum_{k=1}^N \sigma_k^{z}
    \label{eqn:Ising}
\end{equation}
when $j_c < 0$ and $N$ is even. This can be seen from the connection of eq. (\ref{eqn:Ising})
with
eq. (\ref{eqn:isinghamiltonian}) via a Jordan-Wigner transformation \cite{Takahashi,ReuterThesis}. To analyse a translational invariant system with periodic boundary conditions,
we simply change the signs of the second and the fourth terms in the Hamiltonian eq. \eqref{eqn:isinghamiltonian}. 

Now we require only constraints that are due to second moments, i.e. we require that
\begin{equation}
    X=\begin{pmatrix} T & U^{\dagger} \\ U & S \end{pmatrix} \geq 0
\end{equation}
and that the entries fulfil the constrains dictated by the anti-commutator relations.
Solving the relaxed minimization problem eq. \eqref{eqn:relaxedminimizationproblem} by
applying the projected gradient algorithm and comparison of the so-obtained lower bound to
the exact ground state energy \cite{Takahashi,ReuterThesis}
\begin{equation}
    E_{0} =-\sum_{k=0}^{N-1} \left[\left(j_{c}\cos \left(\frac{\pi l_{k}}{N}\right) + h\right)^{2}  + \left(j_{c}\sin\left(\frac{\pi l_{k}}{N}\right)\right)^{2} \right]^{1/2}
\label{eqn:exactIsingenergy}
\end{equation}
yields the results presented in table \ref{tbl:Isingnontranslational}. Note that the ground state energy for the system given by the Hamiltonian eq. \eqref{eqn:isinghamiltonian} is computed by
the formula \eqref{eqn:exactIsingenergy} with parameter $l_{k}=2k+1$, while one chooses $l_{k}=2k$ for the energy of a translational invariant system with periodic boundary conditions.

\begin{table}[htb]
\renewcommand{\arraystretch}{1.5} 
\setlength{\arrayrulewidth}{1pt} 
\centering
\vspace{0.2cm}
\begin{tabular}  {cp{0.2cm}cp{0.2cm}cp{0.2cm}cp{0.2cm}cp{0.2cm}cp{0.2cm}c}
\hline
N & & 1000 & & 5000 & & 10000 & & 50000 & & 100000 & & 500000 \\
\hline
$\Delta E$ & & $2.4\cdot 10^{-8} $ & & $2.5\cdot 10^{-8}$ & & $2.6\cdot 10^{-8}$ & & $2.4\cdot 10^{-8}$ & & $2.4\cdot 10^{-8}$ & & $2.9\cdot 10^{-8}$  \\
\hline
\end{tabular}
\caption{Lower bounds on the ground state energies for the translational invariant Jordan-Wigner transformation of the Ising Hamiltonian with parameters $j_{c}=-1$
and $h=1/2$ for $N$ modes. The Hamiltonian is identical to eq.~\eqref{eqn:isinghamiltonian} but the signs of the second and fourth terms are changed to make the Hamiltonian translational invariant. The relative deviation $\Delta E$ is of the order of the 
parameters of the stopping criteria $\tau_{Dykstra}=\tau_{E}=10^{-8}$, which limits the accuracy of this
method. We initialized the algorithm with $\alpha = N $ for the first step and
$\alpha \ll N$ for the remaining steps until convergence.
}
\label{tbl:Isingnontranslational}
\end{table}

In fact, it can be shown directly that the ground state energy of the Ising model is
{\em exactly} reproduced with the present method as the Ising model can be diagonalized
by an orthogonal mode transformation mapping second moments to second moments, this
making the transformed optimization problem straightforward to solve \cite{Cai}.
Thus the so obtained lower bounds are, in this case, superior to other approaches such
as Anderson lower bounds. In particular, solving the optimization problem for the 
non-translational invariant Hamiltonian eq. \eqref{eqn:isinghamiltonian} for even $N$ and 
negative $j_{c}$ yields the exact ground state energies for the translational invariant Ising 
Hamiltonian eq. \eqref{eqn:Ising}. Even for this Hamiltonian one is able to 
solve the optimization problem for large particle numbers since one is only considering the 
1-positivity conditions which scale linearly in the system size. 

\subsection{Fourth Moments and the Heisenberg model}
While the Ising model is quasi-free it is of considerable interest to study interacting
models, such as the Heisenberg Hamiltonian, that contain, for example, fourth-order fermionic
operators (see \cite{Mazziotti06} for the study of another interacting fermionic system where a different numerical approach is employed). In the 
following, we consider the performance of the proposed algorithm in this situation. The Heisenberg Hamiltonian
\begin{equation}
    H=J \sum_{i=1}^{N-1}\left[ \sigma_{i}^{x}\sigma_{i+1}^{x} + \sigma_{i}^{y}\sigma_{i+1}^{y} + \sigma_{i}^{z}\sigma_{i+1}^{z}\right]
\end{equation}
can be transformed to spinless fermions by the Jordan-Wigner transformation. In this
picture, the Hamiltonian reads
\begin{eqnarray}
    H&=&2J\sum_{i=1}^{N-1} \left[a_{i+1}^{\dagger}a_{i} + a_{i}^{\dagger}a_{i+1}\right] -2J\sum_{i=1}^{N-1}\left[a_{i+1}^{\dagger}a_{i+1} + a_{i}^{\dagger}a_{i}\right] \nonumber\\
    &&\hspace{1cm} +4J\sum_{i=1}^{N-1} a_{i}^{\dagger}a_{i+1}^{\dagger}a_{i+1}a_{i} + J\cdot\left(N-1\right).
    \label{eqn:heisenberghamiltonianaftertransformation}
\end{eqnarray}
Now we are in a position to compare the exact ground state energies of the Heisenberg model with
the lower bounds obtained by applying the suggested algorithm. Indeed, for the Heisenberg Hamiltonian with parameter $J=1/2$ the lower bounds on the
ground state energy lie very close to the actual energies. Table \ref{tbl:Heisenberg} presents
some numerical results.

\begin{table} [htb]
\renewcommand{\arraystretch}{1.5} 
\setlength{\arrayrulewidth}{1pt} 
    \centering
    \vspace{0.2cm}
    \begin{tabular}   { r | p{0.2cm} c p{0.2cm}|p{0.2cm} c p{0.2cm}|p{0.2cm} c p{0.2cm}}
    $N$ 	& & $E_{exact}$ 	&& & Lower bound  & && \textit{Ratio} &\\
    \hline
        6    & &  -4.9872   & & &  -4.9974   & & &  0.9979 & \\ 
        8    & &  -6.7499   & & &  -6.7739   & & &  0.9964 & \\ 
       10   & &  -8.5161   & & &  -8.5557   & & &  0.9954 & \\ 
       12   & &  -10.2842 & & &  -10.3412 & & &  0.9945 & \\ 
       14   & &  -12.0534 & & &  -12.1291 & & &  0.9938 & \\ 
       16   & &  -13.8235 & & &  -13.9189 & & &  0.9931 & \\ 
       18   & &  -15.5940 & & &  -15.7099 & & &  0.9926 & \\ 
       20   & &  -17.3649 & & &  -17.5021 & & &  0.9922 &
    \end{tabular}
\caption{Lower bounds for the ground state energy of the Heisenberg Hamiltonian eq. \eqref{eqn:heisenberghamiltonianaftertransformation}
of $N$ modes and with parameter $J=1/2$. The exact ground state energy is obtained
by diagonalizing the Hamiltonian in the spin picture.
The parameters defining the stopping criteria for Dykstra's algorithm and for the projected gradient
algorithm are $\tau_{Dykstra}=10^{-4}$ and $\tau_{E}=10^{-4}$. Increasing the precision of the algorithm by decreasing these parameters is accompanied by an improvement 
of the lower bounds and an increase of the computing time.}    
\label{tbl:Heisenberg}
\end{table}

\begin{figure}[htb]
\subfigure{
\includegraphics[width=0.40\textwidth]{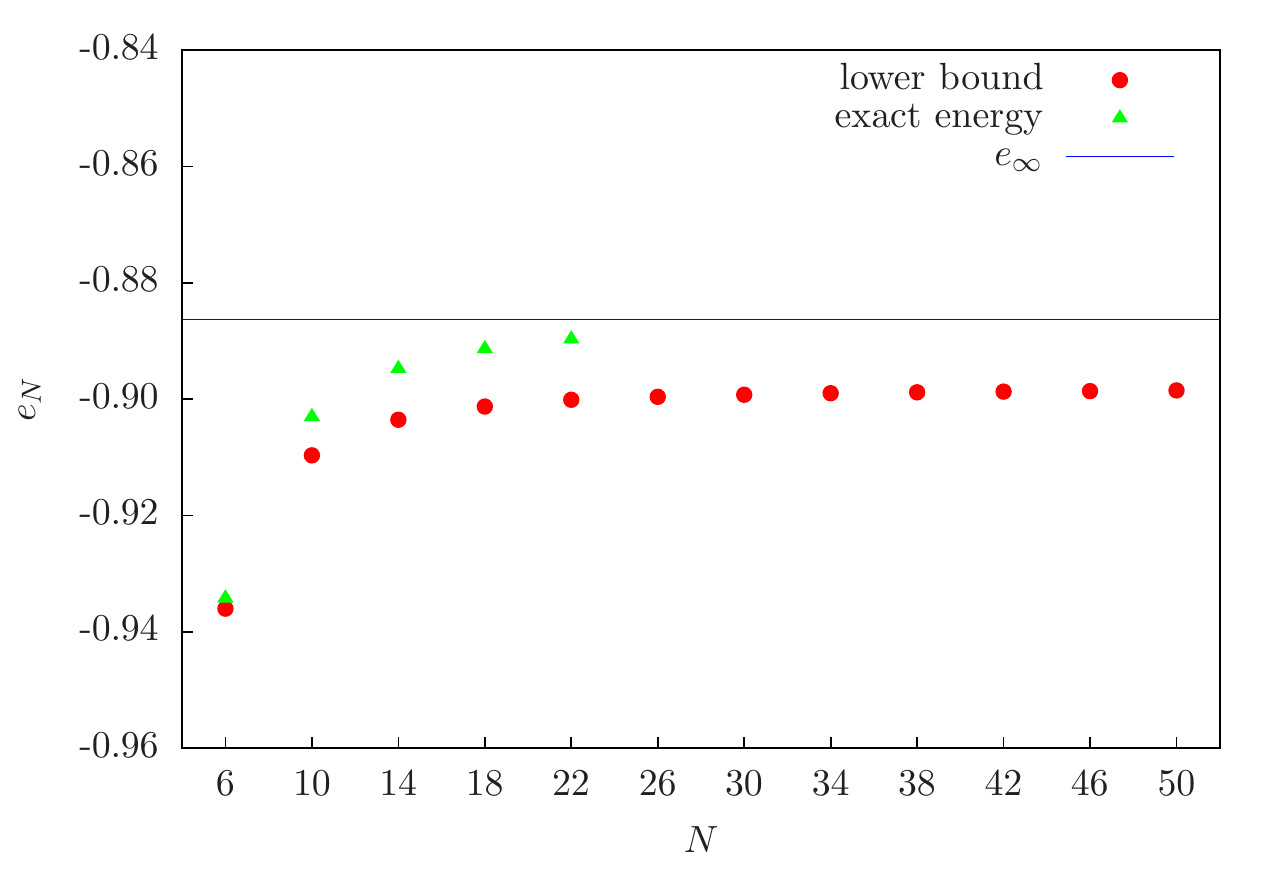}\vspace{-0.7cm}
\qquad
\includegraphics[width=0.40\textwidth]{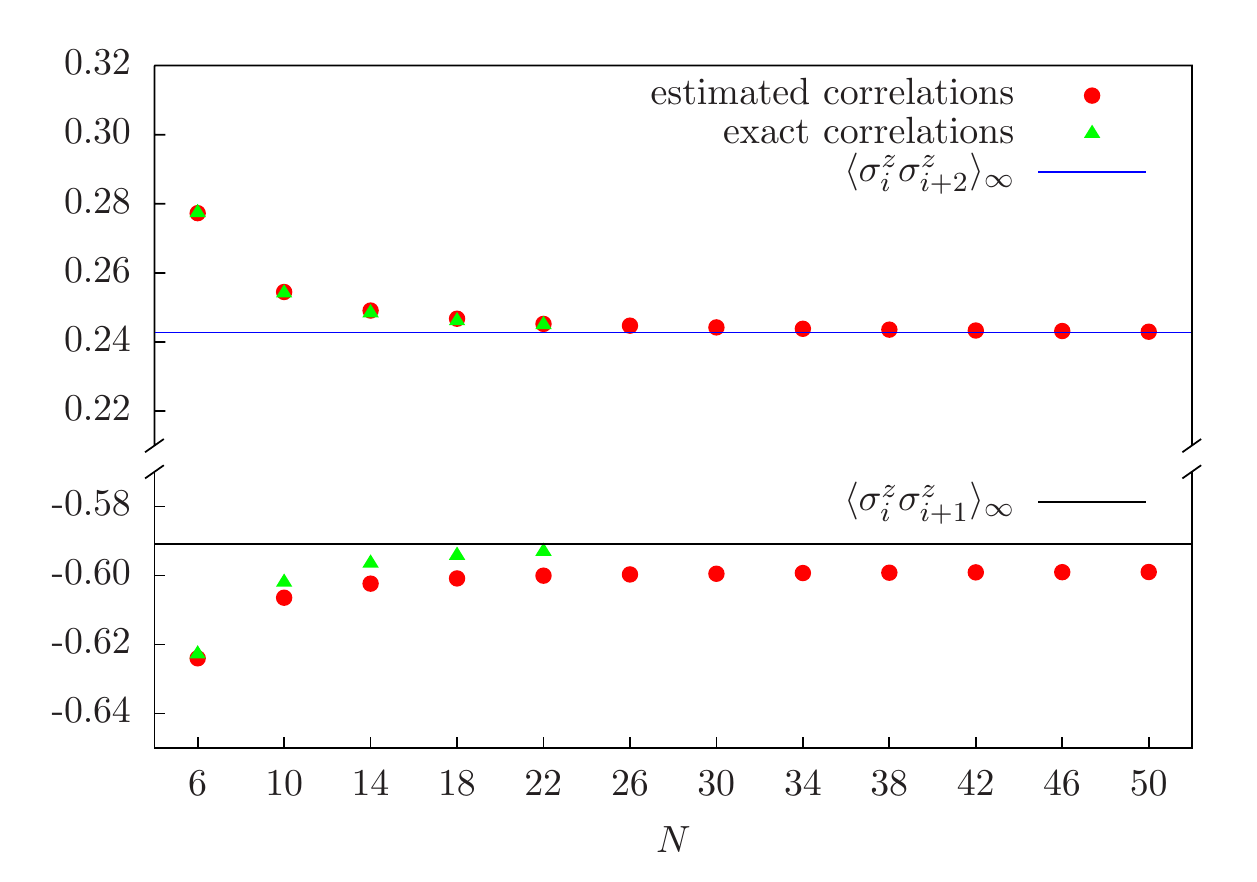}\vspace{-0.7cm}}
\caption{The system under consideration is the Hamiltonian eq.~\eqref{eqn:heisenberghamiltonianaftertransformation} extended to a translational invariant system with periodic boundary conditions and parameter $J=1/2$. The stopping criteria are initialized with $\tau_{Dykstra}=10^{-4}$ and $\tau_{E}=10^{-4}$.
\textit{Left:} lower bounds on the ground state energies $e_{N}=E_{N}/N$ per site. Circles indicate the estimated lower bounds obtained via the projected gradient algorithm. The exact energy
per particle is illustrated as triangles for up to $N=22$ particles. Additionally, the energy per particle in the thermodynamical limit $e_{\infty}= -4|J|(\ln 2 - 0.25)$ \cite{Takahashi} is shown as a horizontal line to indicate where the exact energies converge to.
\textit{Right:}~nearest- and next-nearest-neighbour correlations. Shown are the estimated correlations as circles, the exact correlations as triangles and the correlations in the thermodynamical limit as lines, i.e. ${\langle \sigma_i^z\sigma_{i+1}^z\rangle_{\infty} = (1 - 4\ln 2)/3=-0.59086292}$ and
$\langle \sigma_i^z\sigma_{i+2}^z\rangle_{\infty} = (1 - 16\ln 2 +9\zeta(3))/3 = 0.24271908$
where $\zeta(3)=\frac{5}{2}\sum_{k=1}^{\infty}
(-1)^{k-1}/ [k^3\left(2k \atop k\right) ]\cong 1.2020569...$. The nearest-neighbour correlations are obtained by averaging over
$\langle\sigma_{i}^{\alpha}\sigma_{i+1}^{\alpha}\rangle$ for $\alpha=x,y,z$, which is possible since the system is rotationally invariant. For the next-nearest-neighbour correlations, we proceeded the same way. 
}
\label{fig:energyperparticleHeisenberg}
\end{figure}

The range of applicability of the projected gradient algorithm can be extended
considerably by analysing a system that is translationally invariant. To
this end, we make the Hamiltonian in eq. \eqref{eqn:heisenberghamiltonianaftertransformation} translationally
invariant with periodic boundary conditions. The results for the ground state energy of this Hamiltonian are presented in the left-hand side of figure
\ref{fig:energyperparticleHeisenberg}. As a reference, we depict the exact ground state energies for this model for up to $N=22$. 
Additionally, the right-hand side of figure \ref{fig:energyperparticleHeisenberg} presents the nearest-neighbour and next-nearest-neighbour 
correlations determined by the gradient projection algorithm.

\begin{figure}[htb]
\begin{center}
\includegraphics[width=0.40\textwidth]{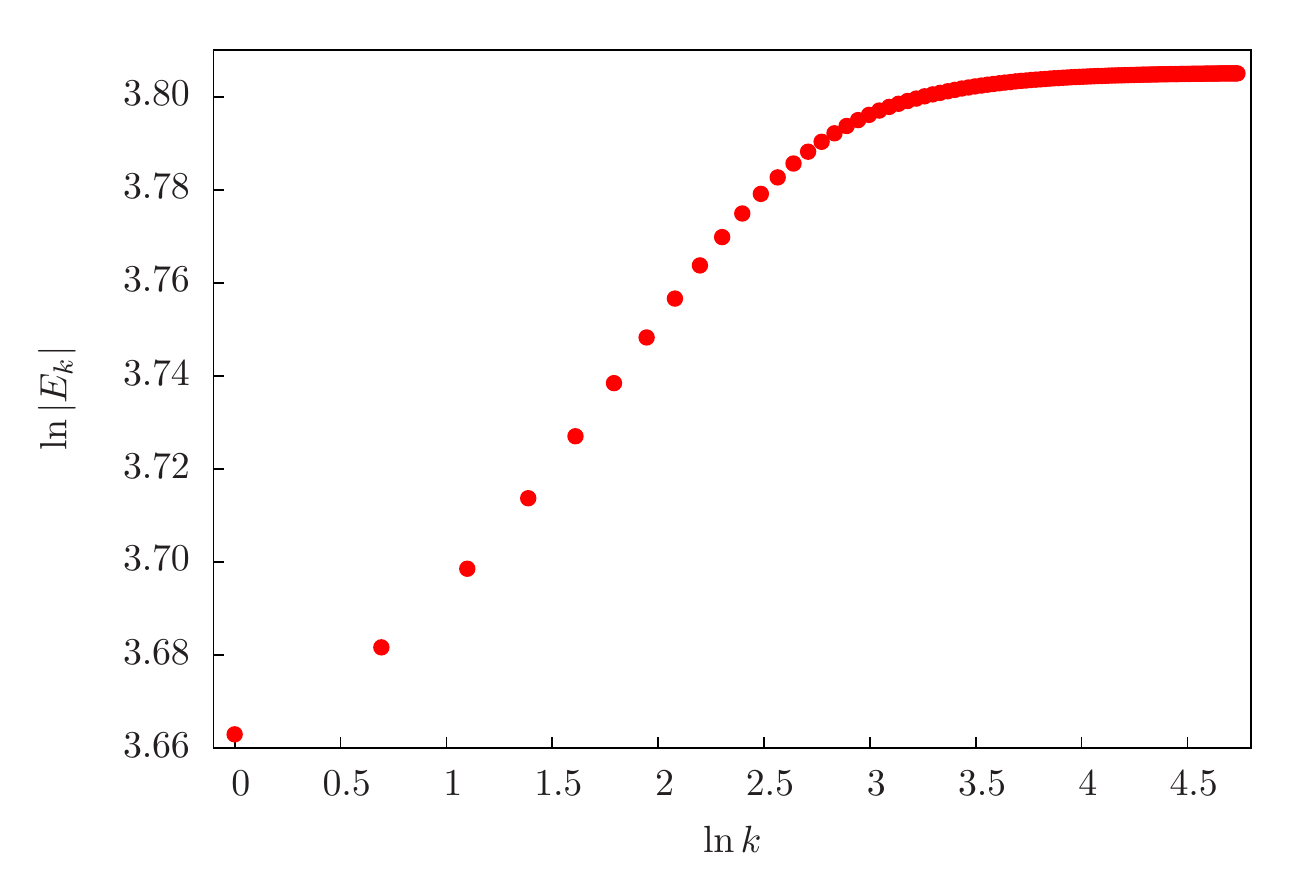}\vspace{-0.7cm}
\end{center}
\caption{Energies $E_{k}$ for each iteration step for the Hamiltonian eq.~\eqref{eqn:heisenberghamiltonianaftertransformation} extended to a translational invariant system for $N=50$ particles illustrating the good 
behaviour of the gradient projection algorithm with respect to the iteration step $k$. Note that the data is plotted double logarithmically. }
\label{fig:rateofconvergenceHeisenberg}
\end{figure}

To illustrate the rate of convergence of the algorithm for fourth moments, we present the
improvement per iteration of the function $E\left(X_{k}\right)=\langle G,X_{k}\rangle +c$ for the translational invariant Jordan-Wigner transformed Heisenberg
Hamiltonian describing $N=50$ sites in figure \ref{fig:rateofconvergenceHeisenberg}. It is
noticeable that the algorithm converges to a good estimation in only a few steps and that the most time-consuming factor is given by the
projection onto the set $\mathcal{C}$. 

Note that in all calculations we initialized the algorithm with the zero matrix since we
wanted to test the algorithm without imposing prior knowledge. Generally, for translational
invariant systems it is advantageous to complete the circulant submatrices obtained
from a run with site number $N_{1}$ to circulant matrices describing $N_{2}$ sites
and use these matrices as an initial matrix for $N_{2}>N_{1}$.

Furthermore, since the matrices comprising the fourth moments in the translational
invariant case decompose into blocks of diagonal matrices, it is possible to parallelize
the projection onto the positive semi-definite matrices. In fact, this step in the computation is
the most expensive since one has to diagonalize the matrices in each step. Parallelizing
this computation might speed up the calculations. So far, no effort has been made to
do so since our intention is rather to demonstrate that this technique yields
good results.

\section{Conclusions}
In this work, we have presented a method for the determination of lower bounds
on the ground state energy of quantum many-body systems. To this end, we followed 
a large body of work to relax the general ground state energy finding problem to 
obtain a semi-definite programme
that involves a number of variables that scale polynomially with the system
size. To overcome unfavourable scaling in time and memory of standard primal-dual
interior point solvers, we combined a projected gradient algorithm with Dykstra's
algorithm to obtain access to larger system sizes and proved its convergence.
For quasi-free systems the method yields the exact ground state energies, while for
general interacting models it provides very good lower bounds and direct access to
correlation functions between sites. Different properties of the underlying system such as 
higher spatial dimensions \cite{mazziotti10}, bosonic degrees of
freedom \cite{mazziotti04_2}, symmetries \cite{mazziotti05} and translational invariance can be included straightforwardly
and the relative weight of higher order correlation functions can be adjusted to
optimize for computational performance.

We hope that further investigations will help to optimize this approach and thus
elucidate its power and limitations.

\section{Acknowledgements}
We acknowledge support from the Bundesministerium f{\"u}r Bildung und Forschung (FK
01BQ1012), the EU Integrating Project Q-ESSENCE and the Alexander von Humboldt
Foundation. Interesting discussions with K. Audenaert, J. Cai, M. Cramer, D.A. Mazziotti and M. 
Navascues are acknowledged. 

\appendix
\section{Proof of Convergence}
\label{app:proofconvergence}
The analysis of lemma \ref{lemma:gradientprojection}  requires some basic properties of the projection operators defined in eq. \eqref{eqn:definitionprojectionontoset}. The following results
can be found, for example, in \cite{deutsch01}, theorems $4.1$ and $5.5$.
\begin{lemma}
Let $P_{\mathcal{C}}:\mathcal{H}\rightarrow \mathcal{C}$ be the projection onto the convex set $\mathcal{C}$.
\begin{enumerate}
\item For all $Z\in\mathcal{C}$ and all $X\in\mathcal{H}$ we have $\langle P_{\mathcal{C}}\left(X\right) - X, Z-P_{\mathcal{C}}\left(X\right) \rangle \geq 0$.
\item For all $X,Y\in\mathcal{H}$ we have $\langle P_{\mathcal{C}}\left(X\right) - P_{\mathcal{C}}\left(Y\right) , X-Y \rangle \geq 0$.
\end{enumerate}
\label{lemma:projectionproperties}
\end{lemma}

The following proof is motivated by \cite{calamai87}.

\textbf{Proof of lemma \ref{lemma:gradientprojection} } \\
Before we start with the main proof, we need to establish the relation
\begin{equation}
\langle G , X_{k}-X_{k+1}\rangle \geq \frac{||X_{k+1} - X_{k}||_{F}^{2}}{\alpha}
\label{equation:1relationmainproof}
\end{equation}
for $\alpha \geq 0$ and where the sequence $\left\{X_{k}\right\}_{k}$ is defined via
\begin{equation}
X_{k+1}=P_{\mathcal{C}}\left(X_{k} - \alpha G\right),
\label{eqn:definitionalgorithm}
\end{equation}
i.e. the projected gradient algorithm. To verify relation \eqref{equation:1relationmainproof}, note that with the first inequality of lemma \ref{lemma:projectionproperties}, we find
\begin{equation}
\begin{split}
&\langle X_{k+1} - X_{k}  + \alpha G , X_{k} - X_{k+1} \rangle \\
&= \langle P_{\mathcal{C}}\left(X_{k} - \alpha G \right) - X_{k}  + \alpha G , X_{k} -P_{\mathcal{C}}\left(X_{k} - \alpha G \right) \rangle \\
&\geq 0
\end{split}
\end{equation}
since $X_{k}\in\mathcal{C}$. Hence we have
\begin{equation}
\langle \alpha G , X_{k} - X_{k+1}\rangle \geq || X_{k} - X_{k+1} ||_{F}^{2}
\end{equation}
and relation \eqref{equation:1relationmainproof} is established. Note that with this expression the following holds for all $\alpha >0$:
\begin{equation}
\langle G , X_{k+1} - X_{k} \rangle \leq 0.
\label{equation:2relationmainproof}
\end{equation}

In a first step, we show that the sequence $E\left(X_{k}\right)$ is monotonically decreasing. We find that 
\begin{equation}
\begin{split}
&\langle G, X_{k+1} \rangle \leq \langle G, X_{k} \rangle, \\
&\Leftrightarrow \langle G, X_{k+1} \rangle + c \leq \langle G, X_{k} \rangle  + c, \\
&\Leftrightarrow E\left(X_{k+1}\right) \leq E\left(X_{k} \right), 
\end{split}
\end{equation}
where $E\left(X\right)= \langle G, X \rangle +c$, which holds for all $\alpha \geq 0$. Hence the sequence $E\left(X_{k}\right)$ is monotonically
decreasing. Since we assume that the function $E\left(X\right)$ is lower bounded on $\mathcal{C}$, we then conclude that this sequence converges for all $\alpha \geq 0$, i.e. $E\left(X_{k}\right)
\rightarrow E^{*}$ for $k\rightarrow \infty$. It remains to be shown that $E^{*}$ is indeed the minimum of the function in $\mathcal{C}$.

In a second step, we show that
\begin{equation}
\limi{k\rightarrow \infty} ||X_{k+1} - X_{k}||_{F} =0.
\end{equation}
Since $E\left(X\right)$ is linear in $X$, we have
\begin{equation}
\begin{split}
E\left(X_{k}\right) &= \langle G, X_{k}\rangle +c +\langle G,X_{k+1} \rangle - \langle G,X_{k+1} \rangle \\
&=\langle G, X_{k+1} \rangle +c + \langle G,X_{k} - X_{k+1}\rangle \\
&=E\left(X_{k+1}\right) + \langle G,X_{k} - X_{k+1} \rangle
\end{split}
\end{equation}
and hence
\begin{equation}
\limi{k\rightarrow \infty} E\left(X_{k}\right) = \limi{k\rightarrow \infty } E\left(X_{k+1}\right) + \limi{k\rightarrow \infty } \langle G, X_{k} - X_{k+1}\rangle.
\end{equation}
Since the sequence $E\left(X_{k}\right)$ converges, i.e. $E\left(X_{k}\right)\rightarrow E^{*}$, we find that 
\begin{equation}
\limi{k\rightarrow \infty } \langle G,X_{k} - X_{k+1}\rangle =0.
\label{equation:skalarproducttozero}
\end{equation}
Inequality \eqref{equation:1relationmainproof} then yields
\begin{equation}
\begin{split}
&\limi{k\rightarrow \infty} \langle G,X_{k} - X_{k+1} \rangle \geq \limi{k\rightarrow \infty } \frac{||X_{k+1} - X_{k}||_{F}^{2}}{\alpha}, \\
&\Leftrightarrow 0 \geq \limi{k\rightarrow \infty } \frac{||X_{k+1} - X_{k}||_{F}^{2}}{\alpha}
\end{split}
\end{equation}
and hence
\begin{equation}
\limi{k\rightarrow \infty } ||X_{k+1} - X_{k}||_{F}^{2} =0.
\end{equation}
Thus we have
\begin{equation}
\limi{k\rightarrow \infty } ||X_{k+1} - X_{k}||_{F} =0.
\label{equation:normdifferencetozero}
\end{equation}

In a third step, we show that the elements $X_{k}$ are bounded. Let $\bar{X}\in\mathcal{C}$ be such that $E\left(\bar{X}\right) \leq E\left(X\right)$ for all
$X\in\mathcal{C}$. Then
\begin{eqnarray}
||X_{k} - \bar{X}||^{2}_{F} &=& ||X_{k} - X_{k+1} + X_{k+1} - \bar{X} ||_{F}^{2} \nonumber\\
& = &||X_{k} - X_{k+1}||_{F}^{2} + ||X_{k+1} - \bar{X}||_{F}^{2} \nonumber\\
&& + 2\langle X_{k} - X_{k+1}, X_{k+1} - \bar{X} \rangle \nonumber\\
& = & ||X_{k} - X_{k+1}||_{F}^{2} + ||X_{k+1} - \bar{X}||_{F}^{2} \nonumber\\
&&+ 2\langle X_{k} -\alpha G + \alpha G - X_{k+1}, X_{k+1} - \bar{X} \rangle \nonumber\\
& = & ||X_{k} - X_{k+1}||_{F}^{2} + ||X_{k+1} - \bar{X}||_{F}^{2} \\
&&+2\langle X_{k} -\alpha G  - P_{\mathcal{C}}\left(X_{k} - \alpha G\right), P_{\mathcal{C}}\left(X_{k} - \alpha G\right) - \bar{X} \rangle \nonumber\\
&&+ 2\langle \alpha G , P_{\mathcal{C}}\left(X_{k} - \alpha G\right) - \bar{X} \rangle \nonumber\\
& = & ||X_{k} - X_{k+1}||_{F}^{2} + ||X_{k+1} - \bar{X}||_{F}^{2} \nonumber\\
&&+2\langle P_{\mathcal{C}}\left(X_{k} - \alpha G\right) - X_{k} +\alpha G , \bar{X} - P_{\mathcal{C}}\left(X_{k} - \alpha G\right) \rangle \nonumber\\
&& + 2\alpha \left[E\left(X_{k+1}\right) - E\left(\bar{X}\right)\right], \nonumber
\end{eqnarray}
since $\langle G , P_{\mathcal{C}}\left(X_{k} - \alpha G\right) - \bar{X} \rangle =\langle G , X_{k+1} - \bar{X} \rangle =E\left(X_{k+1}\right) - E\left(\bar{X}\right)$. Note that with lemma
\ref{lemma:projectionproperties} and the fact that $E\left(\bar{X}\right) \leq E\left(X\right)$ for all $X\in\mathcal{C}$, one can show that the last two terms are greater than or equal to zero.
Therefore
\begin{equation}
||X_{k} - \bar{X}||^{2}_{F} \geq ||X_{k} - X_{k+1}||_{F}^{2} + ||X_{k+1} - \bar{X}||_{F}^{2}
\end{equation}
and consequently
\begin{equation}
||X_{k+1} - \bar{X}||_{F}^{2}  \leq ||X_{k} - \bar{X}||^{2}_{F} -  ||X_{k} - X_{k+1}||_{F}^{2},
\end{equation}
which leads to
\begin{equation}
||X_{k+1} - \bar{X}||_{F}^{2}  \leq ||X_{k} - \bar{X}||^{2}_{F}.
\label{equation:sequencetosolution}
\end{equation}
We conclude that the elements $X_{k}$ are bounded. As a consequence we know that, since $\mathcal{C}$ is closed, there is at least one converging subsequence $X_{n_{k}}\rightarrow X^{*}$
for $k\rightarrow \infty$, where $X^{*}$ is an accumulation point of the sequence $\left\{X_{k}\right\}_{k}$.

In a last step, we show that for any accumulation point $X^{*}$ it holds that $E\left(X^{*}\right) \leq E\left(X\right)$ for all $X\in\mathcal{C}$. With lemma \ref{lemma:projectionproperties}, we
find for all $Z\in\mathcal{C}$
\begin{equation}
\begin{split}
&\langle X_{k+1} - X_{k} + \alpha G , Z - X_{k+1} \rangle \geq 0,  \\
&\Leftrightarrow \alpha \langle G , X_{k+1} - Z \rangle \leq \langle X_{k+1} - X_{k} , Z-X_{k+1} \rangle.
\end{split}
\end{equation}
Further, we have
\begin{equation}
\begin{split}
&\alpha \langle G , X_{k+1} - Z \rangle \\
&\leq \langle  X_{k+1} - X_{k} , Z-X_{k+1} \rangle \\
&\leq \langle X_{k+1} - X_{k} , Z-X_{k} + X_{k} - X_{k+1} \rangle \\
&\leq \langle X_{k+1} - X_{k} , Z - X_{k} \rangle + \langle X_{k+1} - X_{k} , X_{k} - X_{k+1} \rangle \\
&\leq \langle X_{k+1} - X_{k} , Z - X_{k} \rangle - ||X_{k+1} - X_{k}||^{2}_{F} \\
&\leq \langle X_{k+1} - X_{k} , Z - X_{k} \rangle \\
&\leq ||X_{k+1} - X_{k}||_{F} \cdot ||Z-X_{k}||_{F}, 
\end{split}
\end{equation}
where we exploited the Cauchy-Schwarz inequality. Finally, we find that 
\begin{equation}
\begin{split}
\langle G , X_{k} - Z\rangle &= \langle G , X_{k} - X_{k+1}\rangle + \langle G , X_{k+1} - Z\rangle \\
&\leq \langle G , X_{k} - X_{k+1}\rangle + \frac{1}{\alpha} ||X_{k+1} - X_{k}||_{F} \cdot ||Z-X_{k}||_{F} 
\end{split}
\end{equation}
holds for all $Z\in\mathcal{C}$. Now choose a converging subsequence $X_{n_{k}}$ such that $X_{n_{k}}\rightarrow X^{*}$. We already know that ${\langle G,X_{n_{k}} - X_{n_{k}+1} \rangle \rightarrow
 0}$ due to equation \eqref{equation:skalarproducttozero}. Further, equation \eqref{equation:normdifferencetozero} claims that $||X_{n_{k}+1} - X_{n_{k}}||_{F}\rightarrow 0$ for
$k\rightarrow \infty$. Additionally, since the sequence $\left\{X_{k}\right\}_{k}$ is bounded, we know that for $\bar{X}\in\mathcal{C}$, where $E\left(\bar{X}\right)\leq E\left(X\right)$ for all
$X\in\mathcal{C}$,
\begin{equation}
\begin{split}
||Z-X_{n_{k}}||_{F} &= ||Z-\bar{X} + \bar{X} - X_{n_{k}}||_{F} \\
&\leq ||Z-\bar{X}||_{F} + ||\bar{X} - X_{n_{k}}||_{F} \\
&\leq ||Z-\bar{X}||_{F} + ||\bar{X} - X_{0}||_{F} \\
&\leq C\left(Z\right), 
\end{split}
\end{equation}
where we exploited that the sequence $||\bar{X}-X_{n_{k}}||_{F}$ is upper bounded by $||\bar{X} - X_{0}||_{F}$, see equation \eqref{equation:sequencetosolution} , and where $C\left(Z\right)$ is a
constant which may depend on arbitrary $Z\in\mathcal{C}$. Hence we find for all $Z\in \mathcal{C}$ and all accumulation points $X^{*}$ 
\begin{equation}
\begin{split}
&\limi{k\rightarrow \infty} \langle G, X_{n_{k}} - Z\rangle \leq 0, \\
&\Rightarrow \langle G, X^{*} - Z \rangle \leq 0, \\
&\Rightarrow \langle G, X^{*}\rangle \leq \langle G , Z \rangle, \\
&\Rightarrow \langle G, X^{*}\rangle +c \leq \langle G , Z \rangle +c, \\
&\Rightarrow E\left(X^{*}\right) \leq E\left(Z\right).
\end{split}
\end{equation}
Since we know that the sequence $E\left(X_{k}\right)$ converges, we can choose the subsequence defined by
$X_{n_{k}}\rightarrow X^{*}$ to see that $E\left(X_{n_{k}}\right) \rightarrow E^{*}=E\left(X^{*}\right)$. With the argumentation above, we know that $E^{*}\leq E\left(Z\right)$ for all
$Z\in\mathcal{C}$ and hence the algorithm \eqref{eqn:definitionalgorithm} finally ends up at a point minimizing the function $E\left(X\right)$ in the convex set
$\mathcal{C}$. \hfill $\square$

\section{Constraints for a Fermionic System for up to 2-positivity Conditions}
\label{app:constraints}
Here we list the constraints on the matrices for the second and fourth moments which arise
by applying the fermionic anti-commutator relations and further specify the matrices under the additional assumption of particle number 
conservation for a system composed of two spinless fermions. For the second moments of a general, fermionic system, we find
\begin{equation}
\begin{split}
T_{k,l} &=\langle a_{k}^{\dagger} a_{l}\rangle, \\
S_{k,l}&=\langle a_{k}a_{l}^{\dagger} \rangle = \delta_{k,l} - T_{l,k}, \\
U_{k,l}&=\langle a_{k}a_{l} \rangle = -U_{l,k}.
\end{split}
\end{equation}
The fourth moments yield
\begin{eqnarray}
M_{kl,mn}&=&\langle a_{k}^{\dagger}a_{l}^{\dagger} a_{n}a_{m} \rangle =-M_{lk,mn}=-M_{kl,nm}=M_{lk,nm}, \nonumber\\
G_{kl,mn}&=& \langle a_{k}^{\dagger} a_{l}a_{n}a_{m} \rangle = -G_{kl,nm}=G_{km,nl} \nonumber\\
&=&-G_{kn,ml}=-G_{km,ln}=G_{kn,lm}, \nonumber\\
H_{kl,mn}&=& \langle a_{k}a_{l}a_{n}a_{m} \rangle \nonumber\\
&=&-H_{kl,nm}=H_{km,nl}=-H_{km,ln}=H_{kn,lm} \nonumber\\
&=&-H_{kn,ml} =-H_{nl,mk}=H_{nk,ml}=-H_{nm,kl} \nonumber\\
&=&H_{nl,km}=-H_{nk,lm}=H_{nm,lk} =-H_{lk,mn} \nonumber\\
&=&H_{lk,nm}=-H_{lm,nk}=H_{lm,kn}=-H_{ln,km} \\
&=&H_{ln,mk} =-H_{ml,kn}=H_{ml,nk}=-H_{mk,nl} \nonumber\\
&=&H_{mk,ln}=-H_{mn,lk}=H_{mn,kl}, \nonumber\\
R_{kl,mn}&=& \langle a_{k}^{\dagger}a_{l}a_{n}^{\dagger}a_{m} \rangle =\delta_{l,n}T_{k,m} - M_{kn,ml}, \nonumber\\
I_{kl,mn}&=& \langle a_{k}a_{l}a_{n}^{\dagger}a_{m} \rangle =\delta_{l,n}U_{k,m} - \delta_{k,n}U_{l,m} + G_{nk,ml} \mbox{ and } \nonumber\\
Q_{kl,mn}&=& \langle a_{k}a_{l}a_{n}^{\dagger}a_{m}^{\dagger} \rangle\nonumber\\
&=&\delta_{l,n}\delta_{k,m} - \delta_{l,n} T_{m,k} - \delta_{k,n} \delta_{l,m} + \delta_{k,n}T_{m,l} \nonumber\\
&&+ \delta_{l,m} T_{n,k} -\delta_{k,m}T_{n,l}+ M_{nm,lk}.\nonumber
\end{eqnarray}

Assuming particle number conservation for a system composed of two spinless fermions, the matrices containing the second moments read 
\begin{equation*}
T=\begin{pmatrix} \langle a_{1}^{\dagger}a_{1} \rangle & \langle a_{1}^{\dagger} a_{2} \rangle \\ \langle a_{1}^{\dagger} a_{2} \rangle^{*} & \langle a_{2}^{\dagger}a_{2} \rangle \end{pmatrix}, 
\;\;\;\;\;\;\;\;
S=\begin{pmatrix} 1 - \langle a_{1}^{\dagger}a_{1} \rangle & -\langle a_{1}^{\dagger} a_{2} \rangle^{*} \\ -\langle a_{1}^{\dagger} a_{2} \rangle & 1 - \langle a_{2}^{\dagger}a_{2} \rangle \end{pmatrix}, 
\end{equation*}
while the matrices for the fourth moments are 
\begin{equation*}
M =\begin{pmatrix} 0 & 0 & 0 & 0 \\ 0 & \langle a_{1}^{\dagger}a_{2}^{\dagger}a_{2}a_{1} \rangle & -\langle a_{1}^{\dagger}a_{2}^{\dagger}a_{2}a_{1} \rangle  & 0 \\ 
0 & -\langle a_{1}^{\dagger}a_{2}^{\dagger}a_{2}a_{1} \rangle & \langle a_{1}^{\dagger}a_{2}^{\dagger}a_{2}a_{1} \rangle & 0 \\ 0 & 0 & 0 & 0 \end{pmatrix},
\end{equation*}
\begin{equation}
R =\begin{pmatrix} \langle a_{1}^{\dagger}a_{1} \rangle & 0 & \langle a_{1}^{\dagger}a_{2} \rangle & \langle a_{1}^{\dagger}a_{2}^{\dagger}a_{2}a_{1} \rangle  \\ 
0 &  \langle a_{1}^{\dagger}a_{1} \rangle -\langle a_{1}^{\dagger}a_{2}^{\dagger}a_{2}a_{1} \rangle  & 0 & \langle a_{1}^{\dagger}a_{2} \rangle \\ 
\langle a_{1}^{\dagger}a_{2} \rangle^{*} & 0 &  \langle a_{2}^{\dagger}a_{2} \rangle -\langle a_{1}^{\dagger}a_{2}^{\dagger}a_{2}a_{1} \rangle  & 0 \\ 
\langle a_{1}^{\dagger}a_{2}^{\dagger}a_{2}a_{1} \rangle & \langle a_{1}^{\dagger}a_{2} \rangle^{*} & 0 & \langle a_{2}^{\dagger}a_{2} \rangle \end{pmatrix}, 
\end{equation}
\begin{equation*}
Q = \begin{pmatrix} 0 & 0 & 0 & 0 \\
0 & 1 + \langle a_{1}^{\dagger} a_{2}^{\dagger} a_{2}a_{1} \rangle - \langle a_{1}^{\dagger}a_{1} \rangle - \langle a_{2}^{\dagger} a_{2} \rangle &
-1 - \langle a_{1}^{\dagger} a_{2}^{\dagger} a_{2}a_{1} \rangle + \langle a_{1}^{\dagger}a_{1} \rangle + \langle a_{2}^{\dagger} a_{2} \rangle & 0 \\
0 & -1 - \langle a_{1}^{\dagger} a_{2}^{\dagger} a_{2}a_{1} \rangle + \langle a_{1}^{\dagger}a_{1} \rangle + \langle a_{2}^{\dagger} a_{2} \rangle & 
1 + \langle a_{1}^{\dagger} a_{2}^{\dagger} a_{2}a_{1} \rangle - \langle a_{1}^{\dagger}a_{1} \rangle - \langle a_{2}^{\dagger} a_{2} \rangle & 0 \\
0 & 0 & 0 & 0 \end{pmatrix}.
\end{equation*}
The presented matrices $T,S,M,R$ and $Q$ now satisfy the fermionic anti-commutator relations for two spinless fermions. For a system where the particle number is conserved the constraint on the relaxed optimization problem eq. \eqref{eqn:relaxedoptimizationproblem} is that these matrices are all simultaneously positive semi-definite, which generates the 1- and 2-positivity conditions.


\begin{thebibliography}{99}
\bibitem{remarkalgorithm} Indeed, it seems that $\lim_{\alpha\rightarrow \infty} P_{\mathcal{C}}\left(X-\alpha G\right) = X^{*}$ holds, 
where $X\in\mathcal{H}$ is the initial matrix and $X^{*}\in\mathcal{C}$ is such that it minimizes the function $E\left(X\right)$ on the convex set~$\mathcal{C}$, i.e. $E\left(X^{*}\right)\leq
E\left(X\right)$ for all $X\in\mathcal{C}$.
\end{thebibliography}
\end{document}